\documentclass[a4paper]{article}

\usepackage[margin=1in]{geometry}

\usepackage[T1]{fontenc}
\newcommand{\changefont}[3]{
\fontfamily{#1} \fontseries{#2} \fontshape{#3} \selectfont}

\changefont{ptm}{m}{n}

\usepackage{setspace} 
\usepackage{graphicx}    

\usepackage{amsfonts}
\usepackage{amssymb}
\usepackage{graphics}
\usepackage{mathrsfs}
\usepackage{color}

\newtheorem{theorem}{Theorem}[section]

\newtheorem{definition}{Definition}[section]

\long\def\symbolfootnote[#1]#2{\begingroup%
\def\thefootnote{\fnsymbol{footnote}}\footnote[#1]{#2}\endgroup} 

\begin{document}

\begin{center}
\Large \textbf{Existence of Unpredictable Solutions and Chaos}
\end{center}

\begin{center}
\normalsize \textbf{Marat Akhmet$^{a,}\symbolfootnote[1]{Corresponding Author Tel.: +90 312 210 5355,  Fax: +90 312 210 2972, E-mail: marat@metu.edu.tr}$ and Mehmet Onur Fen$^a$} \\
\vspace{0.2cm}
\textit{\textbf{\footnotesize$^a$Department of Mathematics, Middle East Technical University, 06800, Ankara, Turkey\\ $^*$\footnotesize{Corresponding Author Tel.: +90 312 210 5355, Fax: +90 312 210 2972, E-mail: marat@metu.edu.tr}}}
\vspace{0.1cm}
\end{center}

\vspace{0.3cm}

\begin{center}
\textbf{Abstract}
\end{center}

\noindent\ignorespaces

In paper \cite{Arxiv_paper} unpredictable points were introduced based on Poisson stability, and this gives rise to the existence of chaos in the quasi-minimal set. This time, an unpredictable function is determined as an unpredictable point in the Bebutov dynamical system. The existence of an unpredictable solution and consequently chaos of a quasi-linear system of ordinary differential equations are verified. This is the first time that the  description of chaos is initiated from a single function, but not on a collection of them. The results can be easily extended to different types of differential equations. An application of the main theorem for Duffing equations is provided. 

\vspace{0.2cm}
 
\noindent\ignorespaces \textbf{Keywords:} Unpredictable point; Unpredictable function; Poisson stability; Bebutov dynamical system; Quasi-linear differential equation; Unpredictable solution; Chaotic dynamics; Chaos control

\vspace{0.6cm}


\section{Introduction}

The row of  periodic, quasi-periodic, almost periodic, recurrent, Poisson stable motions had been successively developed  in the theory of dynamical systems. Then, chaotic dynamics started to  be considered, which is not a single motion phenomenon, since a prescribed set of motions is required for a definition \cite{li,Devaney}. Our manuscript serves for proceeding the row and involving chaos as a purely functional object in nonlinear dynamics. In our previous paper \cite{Arxiv_paper}, we introduced unpredictable motions based on Poisson stability. This time, we introduce the concept of an unpredictable function as an unpredictable point in the Bebutov dynamics \cite{sell}.

It was proved in \cite{Arxiv_paper} that an unpredictable point gives rise to the existence of chaos in the quasi-minimal set. Thus, if  one shows the  existence  of an unpredictable solution of an equation, then the chaos exists. The result of the present study as well  as our  previous paper concerning  replication of chaos \cite{Akh5} support the opinion of Holmes \cite{holmes90} that the theory of chaos has to be a part of the theory of differential equations. Since the main body of the results on chaotic motions have been formulated in terms of differential and difference equations, we may suggest that all these achievements have to be embedded and developed in the theory of dynamical systems or more specifically, in the theory of differential equations or hybrid systems. 
 
The rest of the paper is organized as follows. In the next section, we give the auxiliary results from the paper \cite{Arxiv_paper}. Section \ref{sec3} is concerned with the Bebutov dynamics and the description of unpredictable functions. The existence of unpredictable solutions in a quasi-linear system is considered in Section \ref{sec4}. Section \ref{sec_example} is devoted to an illustrative example. Finally, some concluding remarks are given in Section \ref{sec_conc}.

\section{Preliminaries} \label{sec_prelim}

Throughout the paper, we will denote by $\mathbb R,$ $\mathbb R_+,$ $\mathbb N$ and $\mathbb Z$ the sets of real numbers, non-negative real numbers, natural numbers and integers, respectively. Moreover, we will make use of the usual Euclidean norm for vectors and the norm induced by the Euclidean norm for square matrices \cite{Horn92}. 
 
Let $(X, d)$ be a metric space. A mapping $\pi: \mathbb R_+ \times X \to X$ is a semi-flow on $X$ \cite{sell} if:
\begin{itemize}
\item[(i)] $\pi(0,p)=p$ for all $p \in X;$
\item[(ii)] $\pi(t,p)$ is continuous in the pair of variables $t$ and $p;$
\item[(iii)] $\pi(t_1, \pi(t_2,p))=\pi(t_1+t_2,p)$ for all $t_1,$ $t_2 \in \mathbb R_+$ and $p \in X.$ 
\end{itemize}

Suppose that $\pi$ is a semi-flow on $X.$ A point $p \in X$ is stable $P^+$ (positively Poisson stable) if there exists a sequence $\left\{ t_n \right\},$ $t_n \to \infty$ as $n \to \infty,$ such that $\pi(t_n,p) \to p$ as $n\to \infty$ \cite{Nemytskii}. For a fixed $p \in X,$ let us denote by $\Theta_p$ the closure of the trajectory $\mathcal{T}(p) = \left\{ \pi(t,p): t \in \mathbb R_+ \right\},$ i.e., $\Theta_p = \overline{\mathcal{T}(p)}.$ The set $\Theta_p$ is a quasi-minimal set if the point $p$ is stable $P^+$ and $\mathcal{T}(p)$ is contained in a compact subset of $X$ \cite{Nemytskii}. 
 
It was demonstrated by Hilmy \cite{Hilmy36} that if the trajectory corresponding to a Poisson stable point $p$ is contained in a compact subset of $X$ and $\Theta_p$ is neither a rest point nor a cycle, then $\Theta_p$ contains an uncountable set of motions everywhere dense and Poisson stable. The following theorem can be proved by adapting the technique given in \cite{Nemytskii,Hilmy36}. 

\begin{theorem} \label{thm1a}   
Suppose that $p \in X$ is stable $P^+$ and $\mathcal{T}(p)$ is contained in a compact subset of $X.$ If \ $\Theta_p$ is neither a rest point nor a cycle, then it contains an uncountable set of motions everywhere dense and stable $P^+.$
\end{theorem}

The results of our paper are correct if one considers stable $P^-$ (negatively Poisson stable) points for a semi-flow with negative time or both stable $P^+$ and stable $P^-$ (Poisson stable) points for a flow. The definition of a quasi-minimal set is given for a Poisson stable point in \cite{Nemytskii}.

The description of an unpredictable point and trajectory are as follows.

\begin{definition} [\cite{Arxiv_paper}] \label{SSP_point}
A point $p \in X$ and the trajectory through it are \textit{unpredictable} if there exist a positive number $\epsilon_0$ (the sensitivity constant) and sequences $\left\{t_n\right\}$  and  $\left\{\tau_n\right\},$ both of which diverge to infinity, such that
  $\displaystyle \lim_{n \to \infty} \pi(t_n,p)=p$  and
  $d[\pi(t_n+\tau_n,p), \pi(\tau_n,p)] \ge \epsilon_0$ for each $n \in \mathbb N.$
\end{definition} 
 
An important point to discuss is the sensitivity or unpredictability. In the famous research studies \cite{li,Devaney,poincare,lorenz,smale}, sensitivity was considered as a property of a system on a certain set of initial data since it compares the behavior of at least couples of solutions. Definition \ref{SSP_point} allows to formulate unpredictability for a single trajectory. Indicating an unpredictable point $p,$ one can make an error by taking a point $\pi(t_n, p).$ Then $d[\pi(\tau_n,\pi(t_n, p)), \pi(\tau_n,p)] \ge \epsilon_0,$ and this is unpredictability for the motion. Thus,  
we say about the unpredictability of a single trajectory whereas the former definitions considered the property in a set of motions.  
  
It was proved in \cite{Arxiv_paper} that if $p\in X$ is an unpredictable point, then $\mathcal{T}(p)$ is neither a rest point nor a cycle, and that if a point $p\in X$ is unpredictable, then every point of the trajectory $\mathcal{T}(p)$ is also unpredictable. It is worth noting that the sensitivity constant $\epsilon_0$ is common for each point on an unpredictable trajectory.

The dynamics on a set $S \subseteq X$ is sensitive \cite{Devaney,lorenz} if there exists a positive number $\epsilon_0$ such that for each $u \in S$ and each positive number $\delta$ there exist a point $u_{\delta} \in S$ and a positive number $\tau_{\delta}$ such that $d[u_{\delta},u]<\delta$ and $d[\pi(\tau_{\delta},u_{\delta}),\pi(\tau_{\delta},u)]\ge \epsilon_0.$
 
A result concerning sensitivity in a quasi-minimal set is given in the next theorem.
  
\begin{theorem} [\cite{Arxiv_paper}]  \label{thm2} 
The dynamics on $\Theta_p$ is sensitive if $p \in X$ is an unpredictable point.
\end{theorem}

Theorem \ref{thm2} mentions the presence of sensitivity in the set $\Theta_p$ if $p$ is an unpredictable point in $X.$ According to Theorem \ref{thm1a}, if the trajectory $\mathcal{T}(p)$ of an unpredictable point $p \in X$ is contained in a compact subset of $X,$ then $\Theta_p$ contains an uncountable set of everywhere dense stable $P^+$ motions. Additionally, since $\mathcal{T}(p)$ is dense in $\Theta_p,$ the transitivity is also valid in the dynamics.  Therefore, summarizing the last discussions, we propose a new definition of chaos  based on the concept  of the   unpredictable point.

\section{Unpredictable Functions and Chaos} \label{sec3}

 This section is devoted to the description of unpredictable functions and their connection with chaos. For that purpose the results provided in \cite{sell} will be utilized.

Let us denote by $C(\mathbb{R})$ the set of continuous functions defined on $\mathbb{R}$ with values in $\mathbb{R}^m,$ and assume that $C(\mathbb{R})$ has the topology of uniform convergence on compact sets, i.e., a sequence $\{h_k \}$ in $C(\mathbb{R})$ is said to converge to a limit $h$ if for every compact set $\mathcal{U}\subset \mathbb{R}$ the sequence of restrictions $\{h_k|_{\mathcal{U}} \}$ converges to $\{h|_{\mathcal{U}} \}$ uniformly.

One can define a metric $d$ on $C(\mathbb{R})$  as \cite{sell}
\begin{eqnarray} \label{distance_formula}
d(h_1,h_2)=\sum_{k=1}^{\infty}{2^{-k}\rho_k(h_1,h_2)},
\end{eqnarray}
where $h_1,$ $h_2$ belong to $C(\mathbb{R})$ and $\rho_k(h_1,h_2)=\min \left\{ 1,\sup_{s\in [-k,k]}\| h_1(s)-h_2(s)\| \right\}, \ k\in \mathbb N.$

Let us define the mapping $\pi:\mathbb{R}_+ \times C(\mathbb{R})\to C(\mathbb{R})$ by $\pi(t,h)=h_t,$ where $h_t(s)=h(t+s).$ The mapping $\pi$ defines a semi-flow on $C(\mathbb{R})$ and it is called the Bebutov dynamical system \cite{sell}.
 	
We describe an unpredictable function as follows. 	
 	
\begin{definition} \label{unpredict_defn}
An unpredictable function	is an unpredictable point of the Bebutov dynamical system.
\end{definition}

According to Theorem $III.3$ \cite{sell}, a motion $\pi(t,h)$ lies in a compact set if $h$ is a bounded and uniformly continuous function. Assuming this, by means of Theorem \ref{thm2}, we obtain that an unpredictable function $h$ determines chaos if it is bounded and uniformly continuous. On the basis of this result, one can say that if a differential equation admits an unpredictable solution which is uniformly continuous and bounded, then chaos is present in the set of solutions. In the next section, we will prove the existence of an unpredictable solution whose quasi-minimal set is a chaotic attractor.
	
\section{Unpredictable Solutions of Quasi-linear Systems}	\label{sec4}
	
Consider the following quasi-linear system,	
\begin{equation}\label{quasi_lin}
x'=Ax+f(x)+g(t), 
\end{equation}	
where the $m \times m$ constant matrix $A$ has eigenvalues all with negative real parts, the function $f: \mathbb R^m \to \mathbb R^m$ is continuous and $g: \mathbb R \to \mathbb R^m$ is a uniformly continuous and bounded function. 	

Since the eigenvalues of the matrix $A$ have  negative real parts, there exist positive numbers $K$ and $\omega$ such that $	\|e^{At}\| \leq K e^{-\omega t},$ $t\geq 0$ \cite{Hale80}.

The following conditions are required.
\begin{itemize}
\item[(C1)] There exists a positive number $M_f$ such that $\displaystyle \sup_{x\in \mathbb{R}^m}\|f(x) \|\leq M_f;$
\item[(C2)] There exists a positive number $L_f$ such that $\left\|f(x_1) - f(x_2)\right\| \le L_f \left\|x_1 - x_2\right\|$ for all $x_1,$ $x_2 \in \mathbb R^m;$
\item[(C3)] $K L_f - \omega <0.$
\end{itemize}	
	
The main result of the present study is mentioned in the next theorem.

\begin{theorem} \label{unpredictable_thm_main}
Suppose that the conditions $(C1)-(C3)$ are valid. If the function $g(t)$ is unpredictable, then system (\ref{quasi_lin}) possesses a unique uniformly exponentially stable unpredictable solution, which is uniformly continuous and bounded on $\mathbb R.$ 
\end{theorem}

\noindent \textbf{Proof.}  Using the technique for quasi-linear equations \cite{Hale80}, one can confirm under the conditions $(C1)-(C3)$ that system (\ref{quasi_lin}) possesses a unique bounded on $\mathbb R$ solution $\phi(t)$ which satisfies the relation 
\begin{eqnarray} \label{integral_eqn}
\phi(t)=\int_{-\infty}^{t} e^{A(t-u)} [f(\phi(u))+g(u)] du.
\end{eqnarray}
Moreover, $\displaystyle \sup_{t \in \mathbb R} \left\|\phi(t)\right\| \le M_{\phi},$ where $M_{\phi}=\displaystyle \frac{K(M_f+M_g)}{\omega}$ and $M_g= \displaystyle \sup_{t\in\mathbb R} \|g(t)\|.$ The solution $\phi(t)$ is uniformly continuous on $\mathbb R$ since $\displaystyle\sup_{t\in \mathbb R} \left\|\phi'(t)\right\|\le \left\|A\right\| M_{\phi} +M_f + M_g.$ 

Suppose that $x(t)$ is a solution of (\ref{quasi_lin}) such that $x(t_0)=x_0$ for some $t_0 \in \mathbb R$ and $x_0\in \mathbb R^m.$ It can be verified that
$$
\left\|x(t) - \phi(t)\right\| \le K \left\|x_0 - \phi(t_0)\right\| e^{(KL_f - \omega)(t-t_0)} , \ t \ge t_0,
$$
and therefore, $\phi(t)$ is uniformly exponentially stable.

Since the function $g(t)$ is unpredictable, there exist a positive number $\epsilon_0 \le 1$ and sequences $\{t_n\},$ $\{\tau_n\},$ both of which diverge to infinity, such that $d(g_{t_n},g)\to 0$ as $n\to \infty$ and $d(g_{t_n+\tau_n},g_{\tau_n}) \geq \epsilon_0$ for all $n \in \mathbb N,$ where the distance function $d$ is given by (\ref{distance_formula}). 

First of all, we shall show that $d(\phi_{t_n},\phi)\to 0$ as $n\to \infty.$ Fix an arbitrary small positive number $\epsilon<1$ and suppose that $\alpha$ is a positive number satisfying $\alpha \le \displaystyle \frac{\omega-KL_f}{2\omega + K - 2 K L_f}.$ Let $k_0$ be a sufficiently large natural number such that 
\begin{eqnarray} \label{proof_ineq1}
k_0 \ge \max \left\{\displaystyle \frac{\ln (1/\alpha \epsilon)}{\ln 2} ,\displaystyle \frac{1}{\omega - KL_f} \ln\left( \frac{2K(M_f+M_g)}{\omega \alpha \epsilon} \right) \right\}.
\end{eqnarray}

There exists a natural number $n_0$ such that if $n \ge n_0$ then $d(g_{t_n},g) < 2^{-2k_0} \alpha \epsilon.$ Therefore, for $n\ge n_0,$ the inequality $\rho_{2k_0} (g_{t_n},g) <\alpha \epsilon$ is valid. Since $\alpha \epsilon<1,$ we have that $\left\|g(t_n+s)-g(s)\right\|<\alpha \epsilon$ for $s\in [-2k_0,2k_0].$

Making use of the relation (\ref{integral_eqn}), one can obtain that 
\begin{eqnarray*}
\phi(t_n+s) - \phi(s) = \displaystyle \int_{-\infty}^s e^{A(s-u)} \left[ f(\phi(t_n+u)) - f(\phi(u)) + g(t_n+u) -g(u) \right] du.
\end{eqnarray*}
Thus, if $s$ belongs to the interval $[-2k_0,2k_0],$ then it can be verified that
\begin{eqnarray} \label{proof_ineq8}
& \left\|\phi(t_n+s) - \phi(s) \right\| & \le  \displaystyle \frac{2K(M_f+M_g)}{\omega} e^{-\omega (s+2k_0)} + \frac{K\alpha \epsilon}{\omega} \left( 1- e^{-\omega (s+2k_0)} \right) \nonumber \\
&& + KL_f \displaystyle \int_{-2 k_0}^s e^{-\omega(s-u)}   \left\| \phi(t_n+u)  -  \phi(u)  \right\|   du.
\end{eqnarray}
Now, let us define the functions $\psi_n(s) = e^{\omega s} \left\| \phi(t_n+s)-\phi(s)\right\|,$ $n \ge n_0.$ Inequality (\ref{proof_ineq8}) implies that
\begin{eqnarray*}
\psi_n(s) \le \displaystyle \frac{K \alpha \epsilon}{\omega} e^{\omega s} + \left( \frac{2K(M_f+M_g)-K\alpha \epsilon}{\omega} \right) e^{-2\omega k_0} + KL_f \displaystyle \int_{-2k_0}^s \psi_n(u) du.
 \end{eqnarray*}
Applying the Gronwall's Lemma \cite{Corduneanu}, one can confirm that 
\begin{eqnarray*}
\psi_n(s) \le \displaystyle \frac{K \alpha \epsilon}{\omega - KL_f} e^{\omega s} \left(1-e^{(KL_f-\omega)(s+2k_0)}\right) + \frac{2K(M_f+M_g)}{\omega} e^{KL_f s} e^{2(KL_f-\omega)k_0}.
\end{eqnarray*}
Hence, the inequality
\begin{eqnarray*}
\left\|\phi(t_n+s) - \phi(s) \right\|  < \displaystyle \frac{K \alpha \epsilon}{\omega - KL_f} + \frac{2K(M_f+M_g)}{\omega} e^{(KL_f-\omega)(s+2k_0)}
\end{eqnarray*}
is valid.
Since the number $k_0$ satisfies (\ref{proof_ineq1}), we have $e^{(KL_f-\omega)k_0} \le \displaystyle \frac{\omega \alpha \epsilon}{2K(M_f+M_g)}$ so that
\begin{eqnarray*}
 \left\|\phi(t_n+s) - \phi(s) \right\|  < \left(1+\frac{K}{\omega-KL_f}\right)\alpha \epsilon, \ s\in[-k_0,k_0].
\end{eqnarray*}
Therefore, the inequality
$$
\sup_{s\in [-k,k]}  \left\|\phi(t_n+s) - \phi(s) \right\| < \left(1+\frac{K}{\omega-KL_f}\right)\alpha \epsilon 
$$ 
holds for each integer $k$ with $1 \le k \le k_0.$ It is clear that $\displaystyle \left(1+\frac{K}{\omega-KL_f}\right)\alpha \epsilon <1.$ Thus, 
$$
\rho_k(\phi_{t_n},\phi) < \displaystyle \left(1+\frac{K}{\omega-KL_f}\right)\alpha \epsilon, \ 1 \le k \le k_0.
$$
For $n \ge n_0,$ it can be obtained by using (\ref{proof_ineq1}) one more time that
\begin{eqnarray*}
& d(\phi_{t_n},\phi) & = \displaystyle \sum_{k=1}^{\infty} 2^{-k}   \rho_k(\phi_{t_n},\phi) \\
&& <  \displaystyle \left(1+\frac{K}{\omega-KL_f}\right)\alpha \epsilon \sum_{k=1}^{k_0} 2^{-k}+ \sum_{k=k_0+1}^{\infty} 2^{-k} \\
&& <   \left(2+\frac{K}{\omega-KL_f}\right)\alpha \epsilon \\
&& \le \epsilon.
\end{eqnarray*}  
Hence,  $ d(\phi_{t_n},\phi) \to 0$ as $n \to \infty.$ 
  
Next, we will verify the presence of a positive number $\overline{\epsilon}_0$ and a sequence $\left\{\widetilde{\tau}_n\right\},$ $\widetilde{\tau}_n \to \infty$ as $n \to \infty,$ such that  $d(\phi_{t_n+\widetilde{\tau}_n},\phi_{\widetilde{\tau}_n}) \geq \overline{\epsilon}_0$ for all $n \in \mathbb N.$
  
Let $N$ be a natural number such that $\displaystyle \sum_{k=N+1}^{\infty} 2^{-k} \leq \frac{\epsilon_0}{2}.$ One can confirm that $$\sum_{k=1}^{N}{2^{-k}\rho_k(g_{t_n+\tau_n},g_{\tau_n})}\geq \frac{\epsilon_0}{2}.$$
In this case, for each $n \in \mathbb N,$ there exist integers $k_0^n$ between $1$ and $N$ such that  
$$
\rho_{k_0^n}(g_{t_n+\tau_n},g_{\tau_n}) \ge \frac{2^{k_0^n}\epsilon_0}{2N}\ge \frac{\epsilon_0}{N}.
$$
Therefore, it can be verified that $$\displaystyle \sup_{s \in [-k_0^n,k_0^n]} \left\|g(t_n+\tau_n+s) - g(\tau_n+s)\right\| \ge \frac{\epsilon_0}{N}, \ n \in \mathbb N.$$ The last inequality implies the existence of numbers $\eta_n \in [-k_0^n,k_0^n]$ satisfying 
\begin{eqnarray} \label{prof_ineq2}
\displaystyle \left\|g(t_n+\tau_n+\eta_n) - g(\tau_n+\eta_n)\right\| \ge \frac{\epsilon_0}{N}, \ n\in\mathbb N.
\end{eqnarray}

Suppose that $g(s)=(g_1(s),g_2(s),\ldots,g_m(s)),$ where each $g_i,$ $1\le i\le m,$ is a real-valued function. In accordance with (\ref{prof_ineq2}), for each $n \in \mathbb N,$ there is an integer $j_n,$ $1\le j_n \le m,$ with
$$
\left|g_{j_n}(t_n+\overline{\tau}_n) - g_{j_n}(\overline{\tau}_n)\right| \ge \displaystyle \frac{\epsilon_0}{Nm},
$$
where $\overline{\tau}_n = \tau_n+\eta_n,$ $n \in \mathbb N.$  
Since the function $g$ is uniformly continuous, there exists a positive number $\Delta \le 1,$ which does not depend on the sequences $\left\{t_n\right\}$ and $\left\{\tau_n\right\},$ such that both of the inequalities
$$
\left\|g(t_n+\overline{\tau}_n)- g(t_n+\overline{\tau}_n+s)\right\| \le \frac{\epsilon_0}{4Nm}
$$
and
$$
\left\|g(\overline{\tau}_n)- g(\overline{\tau}_n+s)\right\| \le \frac{\epsilon_0}{4Nm}
$$
are valid for $s\in [-\Delta,\Delta].$ Thus, we have for $s \in [-\Delta, \Delta]$ that
\begin{eqnarray} \label{proof_ineq4}
& \left|g_{j_n}(t_n+\overline{\tau}_n+s) - g_{j_n}(\overline{\tau}_n+s)\right| & \ge \left|g_{j_n}(t_n+\overline{\tau}_n) - g_{j_n}(\overline{\tau}_n)\right| \nonumber \\ 
&& - \left|g_{j_n}(t_n+\overline{\tau}_n) - g_{j_n}(t_n+\overline{\tau}_n+s)\right| \nonumber \\ 
&& - \left|g_{j_n}(\overline{\tau}_n) - g_{j_n}(\overline{\tau}_n+s)\right| \nonumber \\ 
&& \ge \displaystyle \frac{\epsilon_0}{2Nm}.
\end{eqnarray}
For each $n \in \mathbb N,$ one can find numbers $s_1^n, s_2^n, \ldots, s_m^n \in [-\Delta, \Delta]$ such that
\begin{eqnarray} \label{proof_ineq3}
& &\Big\| \displaystyle \int_{-\Delta}^{\Delta} \left[ g(t_n+\overline{\tau}_n+u) - g(\overline{\tau}_n+u)  \right] du \Big\|  \nonumber \\
&& = 2\Delta \Big( \sum_{i=1}^m  [g_i(t_n+\overline{\tau}_n + s_i^n) - g_i(\overline{\tau}_n+s^n_i)]^2  \Big)^{1/2}.
\end{eqnarray}
Hence, it can be deduced by means of (\ref{proof_ineq4}) and (\ref{proof_ineq3}) that
\begin{eqnarray*}
& \Big\| \displaystyle \int_{-\Delta}^{\Delta} \left[ g(t_n+\overline{\tau}_n+u) - g(\overline{\tau}_n+u)  \right] du \Big\| & \ge 2\Delta \left|g_{j_n}(t_n+\overline{\tau}_n+s^n_{j_n}) - g_{j_n}(\overline{\tau}_n+s^n_{j_n})\right| \\
&& \ge \frac{\Delta \epsilon_0}{Nm}.
\end{eqnarray*}
Now, by using the equation
\begin{eqnarray*}
&\phi(t_n+\bar{\tau}_n+s)-\phi(\overline{\tau}_n+s)& =\phi(t_n+\overline{\tau}_n-\Delta)-\phi(\overline{\tau}_n-\Delta)\\
&& +\displaystyle \int_{-\Delta}^{s} A[\phi(t_n+\overline{\tau}_n+u)-\phi(\overline{\tau}_n+u)] du \\
&& +\displaystyle \int_{-\Delta}^{s} [f(\phi(t_n+\overline{\tau}_n+u))-f(\phi(\overline{\tau}_n+u))] du \\
&& +\displaystyle \int_{-\Delta}^{s} [g(t_n+\overline{\tau}_n+u)-g(\overline{\tau}_n+u)] du, 
\end{eqnarray*}
we attain that 
\begin{eqnarray*}
& \left\|\phi(t_n+\bar{\tau}_n+\Delta)-\phi(\overline{\tau}_n+\Delta)\right\| & \ge \Big\| \displaystyle \int_{-\Delta}^{\Delta} [g(t_n+\overline{\tau}_n+u)-g(\overline{\tau}_n+u)]du 	\Big\| \\
&& - \left\|\phi(t_n+\overline{\tau}_n-\Delta)-\phi(\overline{\tau}_n-\Delta)\right\| \\
&& - \displaystyle \int_{-\Delta}^{\Delta}  \left(\left\|A\right\| + L_f\right)  \left\|  \phi(t_n+ \overline{\tau}_n+u) - \phi(\overline{\tau}_n +u) \right\| du.
\end{eqnarray*}	
The last inequality implies that $$\sup_{s\in[-\Delta,\Delta]} \left\|\phi(t_n+\bar{\tau}_n+s)-\phi(\overline{\tau}_n+s)\right\| \ge \displaystyle \frac{\Delta\epsilon_0}{2Nm [1+\Delta(\left\|A\right\|+L_f)]}, \ n\in \mathbb N.$$	
Therefore, for each $n \in \mathbb N,$ there exists a number $\zeta_n \in [-\Delta,\Delta]$ such that  
$$
\left\|\phi(t_n+\bar{\tau}_n+\zeta_n)-\phi(\overline{\tau}_n+\zeta_n)\right\| \ge \displaystyle \frac{\Delta\epsilon_0}{2Nm [1+\Delta(\left\|A\right\|+L_f)]}.
$$
Let us define the number
$$
\overline{\Delta} = \frac{\Delta\epsilon_0}{8Nm [1+\Delta(\left\|A\right\|+L_f)] [M_{\phi} (\left\|A\right\| + L_f) + M_g]},
$$
and denote $\widetilde{\tau}_n = \overline{\tau}_n+ \zeta_n,$ $n\in \mathbb N.$ Clearly, $\widetilde{\tau}_n \to \infty$ as $n \to \infty.$
One can confirm for $-\overline{\Delta}\le s \le\overline{\Delta}$ that
\begin{eqnarray*}	
& \left\|  \phi(t_n+\widetilde{\tau}_n +s) -\phi(\widetilde{\tau}_n +s) \right\| &	\ge \left\|  \phi(t_n+\widetilde{\tau}_n) -\phi(\widetilde{\tau}_n) \right\| \\
&& - \Big|  \displaystyle \int_0^s   \left( \left\|A \right\| + L_f  \right) \left\|  \phi(t_n+\widetilde{\tau}_n +u) -\phi(\widetilde{\tau}_n +u) \right\| du \Big| \\
&& - \Big|  \displaystyle \int_0^s \left\| g(t_n+\widetilde{\tau}_n +u) - g(\widetilde{\tau}_n +u) \right\| du \Big| \\
&& \ge \displaystyle \frac{\Delta \epsilon_0}{4Nm [1+\Delta (\left\|A\right\|+L_f)]}.
\end{eqnarray*}	
Suppose that $k_1$ is the smallest natural number satisfying $\overline{\Delta} \le k_1.$	In this case, we have for each $k\ge k_1$ that $$\rho_k (\phi_{t_n+\widetilde{\tau}_n}, \phi_{\widetilde{\tau}_n}) \ge \displaystyle \frac{\Delta \epsilon_0}{4Nm [1+\Delta (\left\|A\right\|+L_f)]}.$$ 
Hence, for each $n \in \mathbb N,$ the inequality
\begin{eqnarray*}
d(\phi_{t_n+\widetilde{\tau}_n}, \phi_{\widetilde{\tau}_n}) \ge \displaystyle \sum_{k=k_1}^{\infty} 2^{-k} \rho_k (\phi_{t_n+\widetilde{\tau}_n}, \phi_{\widetilde{\tau}_n}) \ge  \overline{\epsilon}_0,
\end{eqnarray*}	
holds, where
$$
\overline{\epsilon}_0 =   \displaystyle \frac{2^{-k_1+1}\Delta \epsilon_0}{4Nm [1+\Delta (\left\|A\right\|+L_f)]}.
$$

The theorem is proved. $\square$

In the definition of Devaney chaos, periodic motions constitute a dense subset. However, in our case, instead of periodic motions, Poisson stable motions take place in the dynamics.  More precisely, we say that the dynamics on the quasi-minimal set of functions on $\mathbb R$ is chaotic if the dynamics on it is sensitive, transitive and there exists a continuum of Poisson stable trajectories dense in the quasi-minimal set. Nevertheless, in the framework of chaos there may be infinitely many periodic motions. For instance, the symbolic dynamics of bi-infinite sequences possesses both an uncountable set of non-periodic Poisson stable motions as well as infinitely many cycles \cite{Arxiv_paper,Wiggins88}.

\section{An Example} \label{sec_example}
In this part of the paper, we will present an illustrative example. For that purpose, we will make use of coupled Duffing equations such that the first one is forced with a relay function and the second one is perturbed with the solutions of the former. 

Let us consider the following forced Duffing equation,
\begin{equation}\label{ex1}
x''+0.68x'+1.6x+0.008x^{3}=\nu(t,\zeta,\lambda),
\end{equation}
where the forcing term  $\nu(t,\zeta,\lambda)$ is a relay function defined as
\begin{eqnarray} \label{relay}
\nu(t,\zeta,\lambda)=\left\{\begin{array}{ll} 1.2, ~\textrm{if}  & \zeta_{2j}(\lambda) < t  \leq \zeta_{2j+1}(\lambda), ~ j\in \mathbb Z,\\
                                                 0.4, ~\textrm{if}  & \zeta_{2j-1}(\lambda) < t \leq \zeta_{2j}(\lambda), ~ j\in \mathbb Z.
\end{array} \right.
\end{eqnarray}  
In (\ref{relay}), the sequence $\zeta=\left\{\zeta_j\right\}_{j\in \mathbb Z}$ of switching moments is defined through the equation $\zeta_{j}  = j + \kappa_j,$ $j\in \mathbb Z,$ where  the sequence $\left\{\kappa_j\right\}_{j\in \mathbb Z}$ is a solution of the logistic map 
\begin{eqnarray}\label{logistic_map}
\kappa_{j+1} = \lambda \kappa_j (1-\kappa_j).
\end{eqnarray}
By means of the variables $x_1=x$ and $x_2=x',$ equation (\ref{ex1}) can be written as a system in the following form,
\begin{equation}\label{ex2}
\begin{array}{l}
x'_1 = x_2 \\
x'_2 = -1.6 x_1 - 0.68 x_2 - 0.008 x_1^3 + \nu(t,\zeta,\lambda).
\end{array}
\end{equation}

We suppose that the parameter $\lambda$ in (\ref{logistic_map}) is greater than $4$ such that the map possesses an invariant Cantor set $\Lambda \subset [0,1]$ \cite{rob}, and it was demonstrated in \cite{Arxiv_paper} that for such values of the parameter the map possesses an unpredictable point in $\Lambda.$ We consider system (\ref{ex2}) with $\zeta_0 \in \Lambda,$ and the techniques presented in the papers \cite{Akh2}-\cite{Akh3} can be used to prove the existence of a unique unpredictable solution of the system. Moreover, for each natural number $p$, system (\ref{ex2}) admits an unstable periodic solution with period $2p$ if $p$ is odd and an unstable periodic solution with period $p$ if $p$ is even \cite{Akh4}. The reader is referred to \cite{Akh5,Akh2,Akh6,Akh3,Akh4,Akh1} for more information about the dynamics of relay systems.

\begin{figure} 
\centering
\includegraphics[width=11.0cm]{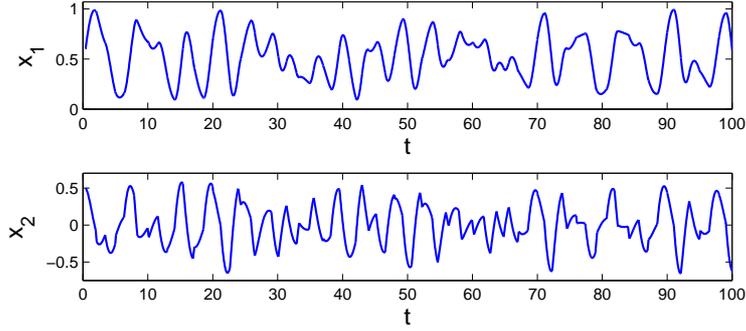}
\caption{The solution of (\ref{ex2}) with $x_1(0.41)=0.6,$ $x_2(0.41)=0.5$ and $\kappa_0=0.41.$ The value $\lambda=4.007$ is used in the simulation. The figure reveals the presence of chaos in the dynamics of (\ref{ex2}).}
\label{fig1}
\end{figure}

In order to demonstrate the chaotic dynamics of (\ref{ex2}), we make use of the value $\lambda=4.007$ in the system and depict the solution corresponding to the initial data $x_1(0.41)=0.6,$ $x_2(0.41)=0.5$ and $\zeta_0=0.41$ in Figure \ref{fig1}. The simulation results seen in Figure \ref{fig1} confirm the presence of chaos in (\ref{ex2}). It is worth noting that even if the logistic map (\ref{logistic_map}) with $\lambda=4.007$ has an invariant Cantor set, the chosen initial value of the sequence $\left\{\zeta_j \right\}$ allows to simulate the solution for $0.41 \le t \le 100.$ Due to the instability, simulations of the system cannot be provided for large intervals of time.

Next, we take into account another Duffing equation,
\begin{equation}\label{ex3}
y''+0.95y'+1.8y+0.005y^{3}=0.
\end{equation}
Using the variables $y_1=y$ and $y_2=y',$ equation (\ref{ex3}) can be reduced to the system
\begin{equation}\label{ex4}
\begin{array}{l}
y'_1 = y_2 \\
y'_2 = -1.8 y_1 - 0.95 y_2 - 0.005 y_1^3.
\end{array}
\end{equation}
We perturb (\ref{ex4}) with the solutions of (\ref{ex2}), and set up the system
\begin{equation}\label{ex5}
\begin{array}{l}
z'_1 = z_2 + x_1(t)\\
z'_2 = -1.8 z_1 - 0.95 z_2 - 0.005 z_1^3 + x_2(t).
\end{array}
\end{equation}
System (\ref{ex5}) is in the form of (\ref{quasi_lin}), where
$A=\left(
\begin {array}{ccc}
0&1\\
\noalign{\medskip}
-1.8&-0.95
\end {array}
\right),$
$f(z_1,z_2)=(0, - 0.005 z_1^3)$ and $g(t)=(x_1(t),x_2(t)).$ Both eigenvalues of $A$ have real parts $-0.475,$ and the coefficient of the non-linear term is chosen sufficiently small in absolute value so that the conditions $(C1)-(C3)$ are valid for (\ref{ex5}). According to Theorem \ref{unpredictable_thm_main}, system (\ref{ex5}) possesses a unique uniformly exponentially stable unpredictable solution, which is uniformly continuous and bounded on $\mathbb R.$ 

\begin{figure} 
\centering
\includegraphics[width=11.0cm]{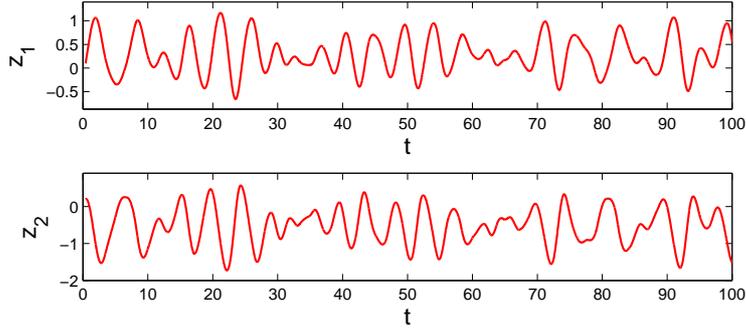}
\caption{Chaotic behavior of system (\ref{ex5}). The solution $(x_1(t), x_2(t))$ represented in Figure \ref{fig1} is utilized as the perturbation in (\ref{ex5}). The irregularity is observable in both $z_1$ and $z_2$ coordinates.}
\label{fig2}
\end{figure}

Figure \ref{fig2} shows the solution of (\ref{ex5}) with $z_1(0.41)=0.1$ and $z_2(0.41)=0.2.$ For the simulation, the solution $(x_1(t), x_2(t))$ which is represented in Figure \ref{fig1} is used. Figure \ref{fig2} supports the result of Theorem \ref{unpredictable_thm_main} such that the represented solution behaves irregularly.

Next, we will demonstrate the presence of periodic motions in system (\ref{ex5}) by means of the Ott-Grebogi-Yorke (OGY) control technique \cite{Ott90}. Since the logistic map (\ref{logistic_map}) is the main source of the chaotic behavior in the coupled system $(\ref{ex2})+(\ref{ex5}),$ we will apply the OGY method to the map. Let us explain  briefly the method for the logistic map \cite{Sch99}. Suppose that the parameter $\lambda$ in (\ref{logistic_map}) is allowed to vary in the range $[4.007-\varepsilon, 4.007+\varepsilon]$, where $\varepsilon$ is a given small positive number. Consider an arbitrary solution $\left\{\kappa_j\right\},$ $\kappa_0\in \Lambda,$ of the map and denote by $\kappa^{(i)},$ $i=1,2,\ldots ,p,$ the target $p-$periodic orbit to be stabilized. 
In the OGY control method \cite{Sch99}, at each iteration step $j$ after the control mechanism is switched on, we consider the logistic map with the parameter value $\lambda=\bar \lambda_j,$ where
\begin{eqnarray}\label{control}
\bar \lambda_j=4.007 \left(1+\frac{(2\kappa^{(i)}-1)(\kappa_{j}-\kappa^{(i)})}{\kappa^{(i)}(1-\kappa^{(i)})} \right),
\end{eqnarray}
provided that the number on the right hand side of the formula $(\ref{control})$ belongs to the interval $[4.007-\varepsilon, 4.007+\varepsilon].$ In other words, formula (\ref{control}) is valid if the trajectory $\left\{\kappa_j\right\}$ is sufficiently close to the target periodic orbit. Otherwise, we take $\bar \lambda_{j}=4.007,$ so that the system evolves at its original parameter value, and  wait until the trajectory $\left\{\kappa_j\right\}$ enters in a sufficiently small neighborhood of the periodic orbit $\kappa^{(i)},$ $i=1,2,\ldots, p,$ such that the inequality $-\varepsilon \le 4.007 \displaystyle\frac{(2\kappa^{(i)}-1)(\kappa_{j}-\kappa^{(i)})}{\kappa^{(i)}(1-\kappa^{(i)})} \le \varepsilon$ holds. If this is the case, the control of chaos is not achieved immediately after switching on the control mechanism. Instead, there is a transition time before the desired periodic orbit is stabilized. The transition time increases if the number $\varepsilon$ decreases \cite{Gon04}.

\begin{figure} 
\centering
\includegraphics[width=11.0cm]{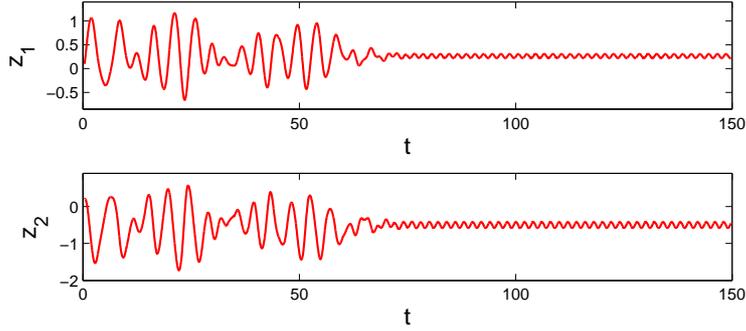}
\caption{The stabilization of the $2-$periodic solution of (\ref{ex5}) corresponding to the fixed point $3.007/4.007$ of the logistic map (\ref{logistic_map}). The value $\varepsilon=0.095$ is used and the control is switched on at $t=\zeta_{20}.$}
\label{fig3}
\end{figure}

Figure \ref{fig3} shows the stabilization of an unstable $2-$periodic solution of (\ref{ex5}).  Here, the OGY control method is used around the fixed point $3.007/4.007$ of the logistic map (\ref{logistic_map}), and the simulation is performed for the initial data $x_1(0.41)=0.6,$  $x_2(0.41)=0.5,$  $z_1(0.41)=0.1,$  $z_2(0.41)=0.2,$ $\zeta_0=0.41.$ The control is switched on at $t=\zeta_{20}$ and the value $\varepsilon=0.095$ is utilized. One can confirm that even if the control is switched on at $t=\zeta_{20}$ there is a transition time before the stabilization such that the control becomes dominant approximately at $t=76.$ Figure \ref{fig3} reveals that the OGY control technique is appropriate for the stabilization of the unstable periodic motions of system (\ref{ex5}).

\begin{figure} 
\centering
\includegraphics[width=11.0cm]{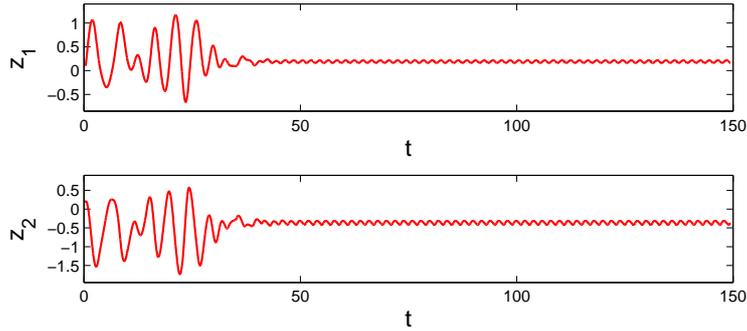}
\caption{The stabilization of the $2-$periodic solution of (\ref{ex5}) corresponding to the $2-$periodic orbit $\kappa^{(1)} \approx 0.34459,$ $\kappa^{(2)}\approx 0.90497$ of (\ref{logistic_map}). The value $\varepsilon=0.072$ is used and the control is switched on at $t=\zeta_{25}.$}
\label{fig4}
\end{figure}

On the other hand, Figure \ref{fig4} shows the simulation result for (\ref{ex5}) when the OGY method is applied around the $2-$periodic orbit $\kappa^{(1)} \approx 0.34459,$ $\kappa^{(2)}\approx 0.90497$ of (\ref{logistic_map}). The represented solution corresponds again to the initial data $x_1(0.41)=0.6,$  $x_2(0.41)=0.5,$  $z_1(0.41)=0.1,$  $z_2(0.41)=0.2,$ $\zeta_0=0.41.$  The value $\varepsilon=0.072$ is used and the control is switched on at $t=\zeta_{25}.$ The presence of a transition time before the stabilization is observable in Figure \ref{fig4} such that the control becomes dominant approximately at $t=46.$ One can observe that the stabilized $2-$periodic solutions seen in Figure \ref{fig3} and Figure \ref{fig4} are different, and this reveals the presence of periodic motions in the quasi-minimal set.

\section{Conclusions} \label{sec_conc}

The unpredictable function as an unpredictable point of the Bebutov dynamics has been defined, and chaos in the quasi-minimal set of the function is verified. This is the first time in the literature that the existence of an unpredictable solution for a quasi-linear ordinary differential equation is proved. Moreover, through simulations it is demonstrated that cycles and non-cyclic Poisson stable orbits can coexist in a quasi-minimal set.

The concept of unpredictable solutions can be useful for finding more delicate features of chaos in systems with complicated dynamics. Researches based on unpredictable functions may pave the way for the functional analysis of chaos to involve the operator theory results. Hopefully, our approach will give a basis for a deeper comprehension and possibility to unite different appearances of chaos. In this framework, the results can be developed for partial differential equations, integro-differential equations, functional differential equations, evolution systems, etc.


\end{document}